\documentclass{article}

\usepackage{amsmath}

\usepackage{amssymb}

\usepackage{cite}

\title{Singularity formation in asymptotically safe cosmology with inhomogeneous equation of state }

\author{Oem Trivedi$^a$ \footnote{oem.t@ahduni.edu.in}  \and Maxim Khlopov \footnote{khlopov@apc.in2p3.fr}  $^{b,c,d}$}
\date{%
	$^a$School of Arts and Sciences, Ahmedabad University,Ahmedabad 380009,India\\%
	$^b$Institute of Physics, Southern Federal University,
	Stachki 194 Rostov on Don 344090, Russia\\%
	$^c$Université Paris Cité, CNRS, Astroparticule et Cosmologie
	F-75013 Paris, France\\%
	$^d$National Research Nuclear University ”MEPHI” 115409 Moscow, Russia\\%
	\today
}

\begin{document}
	
	\maketitle
	
	\begin{abstract}
		Interest in cosmological singularities has remarkably grown in recent times, particularly on future singularities with the discovery of late-time acceleration of the universe and dark energy. While such explorations have previously been done in various modified gravity and quantum gravitationally inspired cosmologies (besides standard general relativistic cosmology), no such an endeavour has been taken up till now in the realms of renormalization group approaches to cosmology and we have hence took up on this journey. In this work, we consider the formation of cosmological singularities in an asymptotically safe cosmology where the cut off scale is proportional to the Hubble parameter. We consider a well motivated inhomogeneous form of the equation of state(EOS) as well. We firstly delve into some basics of this cosmology and show that such a scenario permits a transition between phantom and quintessence forms of universal evolution. We then show that one can have Type I - Type IV singularities in such a cosmology for various version of the EOS and that the singularities can occur both in finite and infinite time. The conditions in which these singularities occur is significantly different than how they occur in the standard cosmology, with the formulations being even more involved. Interestingly this difference comes about without the need of any free parameters in the cosmological models, which is usually what one observes when one goes for the same pursuit via various modified gravity/ QG inspired cosmological approaches. Furthermore, we showed that usual singularity removal methods like conformal anomaly effects and f(R) gravity effects do not offer much hope for singularity removal in this cosmology.
	\end{abstract}
	
	\section{Introduction}
	
	Observations of late time acceleration of the Universe came as a huge surprise to the cosmological community \cite{SupernovaSearchTeam:1998fmf} and ever since then a lot of work has been done in order to explain this expansion. The cosmological expansion problem has been addressed from multiple facets till now, which include the standard approaches of the Cosmological constant \cite{Weinberg:1988cp,Lombriser:2019jia,Padmanabhan:2002ji} alongside more exotic scenarios like Modified gravity theories\cite{Capozziello:2011et,Nojiri:2010wj,Nojiri:2017ncd} and even scalar field driven late-time cosmic acceleration scenarios \cite{Zlatev:1998tr,Tsujikawa:2013fta,Faraoni:2000wk,Gasperini:2001pc,Capozziello:2003tk,Capozziello:2002rd}. Another very interesting way to address cosmological concerns which has picked up pace in recent years is a Renormalization group approach to cosmology, leading to the so called asymptotically safe cosmology.  
	\\
	\\
	While approaches like modified gravity theories affect the whole cosmological dynamics in an extensive fashion, in asymptotically safe cosmology quantum corrections only enforce simple modifications of couplings whose scale dependence gets determined by the renormalization group \cite{Bonanno:2001hi,Bonanno:2001xi,Bentivegna:2003rr,Reuter:2005kb,Bonanno:2007wg,Weinberg:2009wa,Bonanno:2009nj,Bonanno:2010mk,Koch:2010nn,Casadio:2010fw,Bonanno:2010bt,Hindmarsh:2011hx,Cai:2011kd,Contillo:2011ag,Ahn:2011qt,Bonanno:2011yx,frolov2011small}. Asymptotic Safety, in principle, belongs in the same space of theories as the
	corresponding effective field theory. By adding the additional requirement that the quantum theory describing our universe is located within a UV critical hypersurface of any acceptable renormalization group (RG) fixed point, it resolves the predictability problem observed in effective field theory framework. This suggests that the fixed point, which makes all dimensionless coupling constants finite at high energy, controls how the theory behaves at high energies. The running of the cosmological constant and Newton's gravitational constant have been investigated through perturbative RG cosmological techniques, and their implications for big bang nucleosynthesis and other phenomena have been examined recently. They have also shown how there are significant deviations from standard cosmology in such a scenario. Hence asymptotic safety, based on a non-Gaussian fixed point of the gravitational renormalization group flow, can provide an elegant mechanism for completing the gravitational force at sub-Planckian scales.
	\\
	\\
	There has also been an expansive literature in recent times which has been devoted to the study of various types of singularities that could occur during the current and far future of the Universe, with the observation of late-time acceleration having given a significant boost to such works \cite{Nojiri:2004ip,Nojiri:2005sr,Nojiri:2005sx,Bamba:2008ut,trivedi2022finite,trivedi2022type,odintsov2015singular,odintsov2016singular,oikonomou2015singular,nojiri2015singular,odintsov2022did}. These cosmological singularities which have been discussed in recent times can be classified broadly into two types ; strong and weak (such a classification was initially put forward by \cite{ellis1977singular}). Strong singularities are those singularities which can distort finite objects and can mark either the beginning or the end of the universe, with the big bang being the one for the start of the universe and the so called "big rip" signalling the end of the universe. Weak singularities, as the name might suggest, are those which do not have such far reaching implications and do not distort finite objects in the same sense as their strong counterparts. We can discuss these various singularities in more detail as follows : \begin{itemize}
		\item Type I ("Big Rip") : In this case, the scale factor , effective energy density and effective pressure density diverges. This is a scenario of universal death, wherein everything which resides in the universe is progressively torn apart \cite{Caldwell:2003vq}.
		\item Type II ("Sudden/Quiescent singularity") : In this case, the pressure density diverges and so does the derivatives of the scalar factor from the second derivative onwards \cite{Barrow:2004xh}. The weak and strong energy conditions hold for this singularity. Also known as quiescent singularities, but this name originally appeared in contexts related to non-oscillatory singularities \cite{Andersson:2000cv}. A special case of this is the big brake singularity \cite{Gorini:2003wa}.
		\item Type III ("Big Freeze") : In this case, the derivative of the scale factor from the first derivative on wards diverges. These were detected in generalized Chaplygin gas models \cite{bouhmadi2008worse}.
		\item Type IV ("Generalized sudden singularities"): These are finite time singularities with finite density and pressure instead of diverging pressure. In this case, the derivative of the scale factor diverges from a derivative higher than the second \cite{Bamba:2008ut}.
		\item Type V ("w-singularities") : In this case, the scale factor, the energy and pressure densities are all finite but the barotropic index $w = \frac{p}{\rho}$ becomes singular \cite{Fernandez-Jambrina:2010ngm}. 
		\item Type $\infty$(" Directional singularities "): Curvature scalars vanish at the singularity but there are causal geodesics along which the curvature components diverge \cite{Fernandez-Jambrina:2007ohv} and in this sense, the singularity is encountered just for some observers. 
		\item Inaccessible singularities: These singularities appear in cosmological models with toral spatial sections, due to infinite winding of trajectories around the
		tori. For instance, compactifying spatial sections of the de Sitter model to cubic tori. However, these singularities cannot be reached by physically well defined observers and hence this prompts the name inaccessible singularities \cite{mcinnes2007inaccessible}.
	\end{itemize}
	Furthermore, Type I-Type IV singularities were also studied in the context of a cosmology with inhomogeneous equations of state (EOS), firstly in \cite{Nojiri:2005sr,Nojiri:2005sx}. The authors of these papers discussed singularity structures in a cosmology with multiple types of inhomogeneous EOS and showed that Type I-IV singularities can occur in such a cosmology if certain conditions are fulfilled. They also showed how there can be a transition between phantom and quintessence types of evolution for the universes given various inhomogeneous forms for the EOS. These explorations were, however, done in the standard general relativistic cosmology. Hence in this paper we would like to discuss the formation of Type I-IV singularities in an asymptotically safe cosmology with various inhomogeneous equations of state and the paper will be structured as follows \footnote{One can raise the point that asymptotically safe cosmology is only significant in the early universe and not so much for the late universe, where we have considered it. But we would again like to say this is not exactly the case as recent literature to supports the consideration of asymptotically safe cosmology in the very late universe. In particular  it was comprehensively shown in \cite{Anagnostopoulos:2018jdq} as to how asymptotically safe cosmology can be a viable late universe expansion alternative.}. In the next section, we will firstly briefly discuss the basics of asymptotically safe cosmology with an inhomogeneous equation of state after which we will discuss how such a cosmology can transition between phantom and quintessence forms of evolution. In section III we will discuss in detail the singularity structure of this cosmology with three forms of equation of state and we will then conclude our work in section IV.
	
	\section{Brief review of asymptotically safe cosmology with the Phantom-to-Quintessence transition}
	The capacity to build gravitational RG flow approximations outside of perturbation theory is necessary for conceptually testing asymptotic safety. a very strong framework for doing these calculations
	is the functional renormalization group equation (FRGE) for the gravitational
	effective average action $ \Gamma_{k} $ \begin{equation}
	\partial_{k} \Gamma_{k} [g,\overline{g}] = \frac{1}{2} Tr \left[ (\Gamma_{k}^{(2)} + \mathcal{R}_{k} )^{-1} \partial_{k} \mathcal{R}_{k} \right]
	\end{equation} 
	The construction of the FRGE uses the background field formalism, where the
	metric $ g_{\mu \nu} $ is split into a fixed background $ \overline{g}_{\mu \nu} $ and fluctuations $ h_{\mu \nu} $. The Hessian $\Gamma_{k}^{(2)}$
	is the second functional derivative of $ \Gamma_{k} $ with respect to the fluctuation field in a
	fixed background and a scale-dependent mass term for fluctuations with momenta $ p^2 << k^2 $ (where the RG scale k constructed from the background
	metric) is provided by $ \mathcal{R}_{k} $. As it is a mass term, we can assume it to be a positive definite in nature. The role that  $\mathcal{R}_{k}$ plays in the numerator and denominator gives us a trace which could be both infrared and ultraviolet finite and also ensures that the flow of $\Gamma_{k}$ is
	actually under the jurisdiction of the fluctuations with momentum $ p^2 \approx k^2 $ . 
	A useful approach towards asymptotic safety is the functional renormalization group equation for the gravitational effective average action $\Gamma_{k} $.
	\\
	\\
	One could argue that the simplest approximation of the gravitational RG flow could be obtained from projecting the FRGE onto the Einstein-Hilbert action approximating $ \Gamma_{k}$ by \cite{Bonanno:2017pkg} \begin{equation}
	\Gamma_{k} = \frac{1}{16 \pi G_{k}} \int d^4 x \sqrt{-g} \left[-R + 2 \Lambda_{k}\right] + \text{gauge-fixing and ghost terms}
	\end{equation}
	where R, $ \Lambda_{k} $ and $ G_{k} $ are the Ricci Scalar, the running cosmological constant and the running Newton's gravitational constant. The scale-dependence of these couplings can be written in terms of their dimensionless counterparts as \begin{equation}
	\Lambda_{k} = k^2 \lambda_{*}
	\end{equation}
	\begin{equation}
	G_{k} = g_{*}/k^{2}
	\end{equation}
	where $g_{*} = 0.707$ and $ \lambda_{*} = 0.193$ .
	There has been a significant literature which has been invested in considering various forms of the cutoff scale k \cite{Bonanno:2017pkg} . In order to move forward, we consider the background metric to be that of a flat FLRW cosmology \begin{equation}
	ds^2 = -dt^2 + {a(t)}^2 (dx^2 + dy^2 + dz^2 )
	\end{equation}
	Then, considering a perfect fluid form for the stress-energy tensor , $ T_{\mu}^{\nu} = \text{diag} [-\rho,p,p,p] $ , one can get the Friedmann equation and the continuity equation in this scenario to be \begin{equation}
	H^{2} = \frac{8 \pi G_{k}}{3} + \frac{\Lambda_{k}}{3}
	\end{equation} 
	\begin{equation}
	\dot{\rho} + 3 H(\rho + p) = - \frac{\dot{\Lambda_{k}} + 8 \pi \dot{G_{k}}}{8 \pi G}
	\end{equation}
	Where the continuity equation comes about from the Bianchi identity which is satisfied by Einstein's equations $ D^{\mu} [\lambda(t) g_{\mu \nu} - 8 \pi G(t) T_{\mu \nu} ] = 0 $, which has the usual meaning that the divergence $ D^{\mu}$ of the Einstein tensor vanishes . The extra terms of the right hand side in (2.7) can be interpreted as an illustration of the energy transfer between gravitational degrees of freedom and matter.  
	To proceed further, we need a form for the cut-off scale k. In the RG improvement scheme, the cutoff scale k is usually associated with a length scale for the system. In the context of cosmology, one could have severeal interesting identifications (as discussed in \cite{Bonanno:2017pkg} for example) A popular form for k where one uses k to be proportional to the Hubble parameter, where the Hubble parameter is a function of time (which represents a proper distance and hence this could be a viable length scale identification for the system) and so here we take k as \footnote{One might be tempted to think that this cut-off identification is arbitrary but it is hardly so.The cut-off choice can indeed be very different from this as well Taking the cut-off scale proportional to the Hubble parameter has been a recurring theme in asymptotically safe cosmology studies and one can see for example \cite{Bonanno:2017pkg} , \cite{mandal2020cosmology} , \cite{Bonanno:2011yx} to look at discussions about this choice of scale. In particular one can read the discussion in \cite{Bonanno:2017pkg}, for instance, which discusses four different types of cut-off identifications with the one where the cut-off choice is proportional to the Hubble parameter being one of them. This choice was firstly introduced in \cite{Bonanno:2007wg} and there has been significant work on this basis ever since.} \begin{equation}
	k^2 = \epsilon {H(t)}^2 
	\end{equation}
	where $\epsilon$ is an a priori undetermined positive parameter of $\mathcal{O}(1)$. Using this cutoff scale and the values (2.3-2.4), we can write the Friedmann and continuity equations as \begin{equation}
	H^4 \approx \frac{18 \rho}{3 \epsilon - 0.193 \epsilon^2 } = M \rho
	\end{equation} 
	\begin{equation}
	\dot{\rho} \left( \frac{1}{2 \rho} - 1 \right) + 3 H (\rho + p) \approx 0 
	\end{equation}
	where $ M = \frac{54}{ \epsilon - 0.064 \epsilon^2 } $ . The form of the equation of state we will be considering here is given by \begin{equation}
	p = - \rho - f(\rho)
	\end{equation} This sort of equation of state with $ f(\rho)  = A \rho^{\alpha} $ for $\alpha$ being an arbitrary constant was first proposed in \cite{Nojiri:2004pf} and was investigated in detail in \cite{Stefancic:2004kb} and there can be diverse physical motivations behind such an equation of state. This form of EOS can also be equivalent to bulk viscosity \cite{Barrow:1986yf}. This type of an equation of state can also come about due to modified gravity effects \cite{Nojiri:2005sr}.  In order to now discuss a model for transition between phantom and quintessence forms of evolution we consider the following ansatz for the Hubble parameter \cite{Nojiri:2005sx} \begin{equation}
	H = n \left( \frac{1}{t} + \frac{1}{t_{s} - t} \right)
	\end{equation}
	where $ t_{s}$ refers to the time of the singularity and in this case, it can also be taken as the age of the universe as when $ t = t_{s} $, we see a big rip scenario \footnote{The ansatz above really comes about by the consideration that the scale factor is given by $ a(t) = a_{o} \left( \frac{t}{t_{s} - t} \right) $ So the real starting point here is the scale factor but we have proceeded from the Hubble factor right away} and n is a real constant. Using the Friedmann equation (2.9), the energy density can be written as \begin{equation}
	\rho = \frac{ n^{4} {t_{s}}^{4} }{M (t_{s} -t )^4 t^4 }
	\end{equation}
	And using the continuity equation (2.10) and equation of state (2.11), we can get \begin{equation}
	f(\rho) = \pm \frac{2 (1 - 2 \rho) }{3 n t_{s}} \sqrt{t_{s} \left( t_{s} \pm \frac{ 4 n}{M^{1/4} \rho^{1/4} } \right)}
	\end{equation}
	where $f(\rho) $ turns out to be a 4-valued function instead of a double valued function as it happens in standard cosmology with the same ansatz \cite{Nojiri:2005sx} and we start to see the effects of asymptotic safety on cosmological dynamics here. The equation of state in this scenario is given by \begin{equation}
	w = -1 \pm \frac{2 (1 - 2 \rho) }{3 n t_{s} \rho} \sqrt{t_{s} \left( t_{s} \pm \frac{4 n}{M^{1/4} \rho^{1/4} } \right)}
	\end{equation}
	Although the function $f(\rho) $ is technically four valued, the $ \pm $ term in the square roots is not that important in the context of the equation of state, as we want to see how w can move away from -1 (also we can just take the plus part of the square root function if we want, without any loss of generality). We see that w in this case can transition from phantom region $ w < -1 $ to the quintessence region $ w> -1 $ depending on the parametrization of $f (\rho) $, again as in  \cite{Nojiri:2005sx} . This kind of behaviour of the equation of state is similar to what happens for first order phase transitions and hence, in an asymptotically safe cosmology with such inhomogeneous equations of state we can have transitions between phantom and quintessence evolution of the universe.
	
	\section{Singularity formation for various forms of equation of state}
	In this section, we will discuss singularity formation in the asymptotically safe cosmology in detail.First up, we now express Friedman equation (2.9) as \begin{equation}
	H^4 = M \rho
	\end{equation}
	where  $ M = N/3 $. While this might seem insignificant, we feel that re parametrizing like this will be more convenient for the forthcoming formulations. Using the Friedman equation as above and the continuity equation (2.10) and assuming the equation of state (2.11), we can find the scale factor to be \begin{equation}
	a = a_{o} \exp \left[ \int \frac{1 - 2 \rho}{6 \rho f(\rho)} d \rho  \right]
	\end{equation}
	And furthermore, we can write a relation between the cosmological time t and $\rho$ as \begin{equation}
	t - t_{s} =  \int \frac{1 - 2 \rho}{6 M^{1/4} \rho^{5/4} f(\rho)} d \rho 
	\end{equation}
	The above two equations will allow us to identify about different types of cosmic singularities that can occur in this scenario. Broadly, we can write that \begin{itemize}
		\item For Type I singularities $ t \to t_{s} $ , $ a \to \infty $, $ p \to \infty $ and $ \rho \to \infty $ . 
		\item For Type II singularities $ t \to t_{s} $ , $ a \to a_{s} $, $ p \to \infty $ and $ \rho \to \rho_{s} $ or $  0 $ .
		\item For Type III singularities $ t \to t_{s} $ , $ a \to a_{s} $, $ p \to \infty $ and $ \rho \to \infty $ . 
		\item For Type IV singularities $ t \to t_{s} $ , $ a \to a_{s} $, $ p \to p_{s} $ and $ \rho \to  \rho_{s} $ but derivatives of the Hubble Factor H diverge from the second derivative onward. 
	\end{itemize}
	With this cleared up , we would now like to consider two different forms of $ f(\rho) $ and discuss singularity formations for each of these forms. 
	
	\subsection{$f(\rho) = A \rho^{\alpha} $ }
	Here we will consider the case of $ f(\rho) = A \rho^{\alpha} $ , for $\alpha$ being an arbitrary constant and A is some real number. For this case (2.10) becomes \begin{equation}
	\dot{\rho} \left( \frac{1}{2 \rho} - 1 \right) = 3 H (A \rho^{\alpha}) 
	\end{equation}
	Using (3.1), we can further write this as \begin{equation}
	\dot{\rho} \left( \frac{1}{2 \rho} - 1 \right)  = 3 M^{1/4}A \rho^{1/4 + \alpha}
	\end{equation}
	We can now use (2.11) to find the scale factor as a function of the energy density, which comes out to be \begin{equation}
	a(\rho) = a_{o} \exp \left[\frac{\rho^{-\alpha} \left(1 + \alpha (2 \rho - 1 ) \right)  }{6 A \alpha (\alpha - 1 )}\right]
	\end{equation}
	We can further write a relation between $\rho$ and cosmic time t (for $ \alpha \neq \frac{3}{4} - \frac{1}{4} $) \begin{equation}
	t - t_{o} = \frac{2 \rho^{-\alpha -1/4} (3-4 \alpha + 2 \rho + 8 \alpha \rho) }{3 A M^{1/4} (4 \alpha -3) (1+4 \alpha) } 
	\end{equation}
	while for $\alpha = \frac{3}{4}$ we have \begin{equation}
	t - t_{o} = \frac{1}{6 A M^{1/4} } \left[ \frac{\rho - \rho_{o} }{2 \rho \rho_{o}} - 2 \ln \left( \frac{\rho}{\rho_{o}} \right)\right]
	\end{equation} 
	(where $ \rho_{o} $ is the energy density at time $t_{o}$) 
	and for $\alpha = - \frac{1}{4}$ we have \begin{equation}
	t - t_{o} = \frac{1}{3 M^{1/4} A} \left[ \frac{1}{2} \ln \left( \frac{\rho}{\rho_{o}}\right)-  (\rho - \rho_{o})\right]
	\end{equation}
	Now, we can immediately compare these formulas with their corresponding forms in the usual standard cosmology with the scale factor being \begin{equation}
	a(\rho) = a_{o} \exp \left[ \frac{\rho^{1-\alpha}}{3A (1-\alpha)}  \right]
	\end{equation}
	and the relation between t and $\rho$ for $\alpha \neq \frac{1}{2}$ being \begin{equation}
	t - t_{o} = \frac{k \rho^{-\alpha + 1/2} }{A (1-2\alpha)}
	\end{equation}
	where k is a constant and for $\alpha = 1/2 $ \begin{equation}
	t-t_{o} = \frac{k}{2A} \ln \left( \frac{\rho}{\rho_{o}}\right)
	\end{equation}
	We immediately see some very significant changes between the formulations in usual cosmology and the one in asymptotically safe cosmology and hence, there will be quite a few significant differences in the inferences drawn from (3.10-3.12), as was the case for  (3.6-3.9) in \cite{Nojiri:2005sr}.For example, in the usual cosmology for $ \alpha = 1/2$, $\rho$ can only diverge in the far future but in this case, for $\alpha = 1/2$ we can have a finite time divergence of $\rho$ as well. In the usual cosmology the cases $ \alpha = 3/4 $ and $\alpha = -1/4 $ did not have any special solutions but here they have special solutions. 
	\\
	We would now like to discuss the singularity structure in detail. Firstly, the scale factor here takes a very interesting form and we can draw some useful conclusions from this.  \begin{itemize}
		\item When $ \alpha > 1 $ then the scale factor will always be finite for $ \rho \to \infty $ regardless of whether $ A > 0 $ or $ A < 0 $. As $ \rho \to \infty $, $ p \to \infty $ because of the EOS (2.11) and so for $\alpha > 1$ we will have type III singularities.
		\item For $\alpha < 1 $ and $ A > 0 $ , $ a \to 0 $ as $\rho \to \infty$ and so in this case, we will again be dealing with a type III singularity. \footnote{ One might be tempted to think that for a null scale factor we could be dealing with a big bang singularity in this case, but that is not true as the equation of state parameter $ w = -1 $. This can, however, be considered as a possible grand bang/rip \cite{fernandez2014grand} but we will not be considering the grand bang/rip singularities in our paper}
		\item For $\alpha < 1 $ and $ A < 0 $ , $ a \to \infty $ as $\rho \to \infty$ and so in this case as all of a, $\rho$ and p diverge, we will have type I or Big rip singularity.
	\end{itemize}
	Furthermore, we can now discuss the special cases of the $\rho - t $ relation (3.8-3.9). Firstly as $\rho \to \infty$ the right hand side of both (3.8) and (3.9) diverge towards negative infinity. We can now comment for the cases $ \alpha \neq \frac{3}{4} - \frac{1}{4} $ as follows \begin{itemize}
		\item For $ \rho \to \infty $ and $ A > 0 $, $ t \to - \infty $ and as $\rho$ diverges, so will the pressure as well. In this case, the scale factor will still stay finite (3.6). Hence, this points towards the fact that for these values of $\alpha$ , we have a Type III singularity in the far past for a positive value of the parameter A. 
		\item For $ \rho \to \infty $ and $ A < 0 $, $ t \to  \infty $ and the pressure p diverges again, but in this case the scale factor a will also diverge as $\alpha < 1$ as well. So we see that if $\alpha$ takes these values for $ A < 0 $, then we can have a big rip (Type I) singularity in the far future of the universe \footnote{A reader could also be tempted by the possibility of a logarithmic singularity in (3.8-3.9) for $\rho \to 0 $ and we would say that in principle that is a viable possibility but we have not considered that here as the possibility of vanishing energy densities in the far future of the universe is something which is not in the current scope of the work we would want to consider here. }.
	\end{itemize}
	We can now discuss the singularity formations for the more general case (3.7) in more detail \begin{itemize}
		\item For the case when $\alpha > 1$ , when $ \rho \to \infty $ then p will also diverge (2.11) first of all. Secondly, the right hand side of (3.7) will tend towards zero and hence $ t \to t_{o} $ and this would be the result regardless of whether A is positive or negative. The scale factor does not diverges in this scenario as $ \alpha > 1 $. Hence, in this case we can have a type III singularity in finite time.
		\item For the case when $3/4 < \alpha < 1$ , when $ \rho \to \infty $ then p also diverges. The right hand side of (3.7) will again tend towards zero and hence $ t \to t_{o} $ and this would again be the result regardless of whether A is positive or negative. Through (3.6), a can still diverge in this scenario as $ \alpha < 1 $ for $ A < 0 $. Hence, in this scenario we can have a type I singularity in finite time. 
		\item For the case when $-1/4 < \alpha < 3/4$ , when $ \rho \to \infty $ then p will again diverge. This is the first case where we the right hand side of (3.7) diverges and hence the singularities occuring in this case will occur in either the infinite future or the infinite past. For $ A > 0 $, $ t \to - \infty $ and as the scale factor will not diverge in this case (3.6), we have a Type III singularity in the far past. For $ A < 0 $, $ t \to \infty $ and as the scale factor diverges in this case, we have a type I singularity in the far future in this scenario.
		\item For the case when $ \alpha < -1/4 $, when $\rho$ diverges the right hand side of (3.7) diverges as well. In this scenario when $ A> 0 $, $ t \to -\infty $ , and as for positive A we can only have a finite scale factor, in this case the only the pressure and energy densities diverge in the far past and hence we have a type III singularity in this scenario. For $ A < 0 $, $ t \to \infty $ and as in this case the scale factor also diverges ( as $ A < 1 $)  we have a type I singularity in the far future in this case. 
		\item Finally, we would also like to comment on the possiblity of Type IV singularities. We can write Hubble factor H as \begin{equation}
		H = \frac{\dot{a}}{a} = \frac{da}{d\rho} \dot{\rho}
		\end{equation}
		By using the continuity equation (3.5) and the scale factor (3.6), we can write the Hubble factor as \begin{equation}
		H = \frac{M^{1/4} \rho^{1/4}}{A}
		\end{equation}
		Its very interesting to note that the Hubble parameter does not depend on $\alpha$ in any way and the hubble parameter will diverge for $\rho \to \infty$.Its second derivative,however, does depend on $\alpha$ and is given by \begin{equation}
		\ddot{H} = \frac{9 M^{3/4} \rho^{2 \alpha+\frac{3}{4}} (\alpha (2-4 \rho)+2 \rho+1)}{2 A (1-2 \rho)^3} 
		\end{equation} 
		The second derivative of H remains finite for $ \alpha > 0 $ (regardless of whether A is positive or negative) when $ \rho \to 0 $ , $ p \to 0 $. However, the second derivative (and the subsequent derivatives ) diverge  for $\alpha < 0$ (for both positive and negative values of A) but in this case even though $\rho \to 0$, p diverges. Hence, we cannot have a Type IV singularity with this form of the EOS for $ \rho \to 0 $ . However, if $ \rho \to 1/2 $ then we although the second derivative of H diverges, neither the scale factor nor the energy or pressure densities diverge. Hence, we can have a type IV singularity in this case for $ \rho \to 1/2 $. Furthermore looking at the expressions (3.7-3.9), we see that for $ \rho \to 1/2$, the cosmic time t does not diverge unless $\alpha $ values are tuned for it diverge. So we can have type IV singularities in a finite timeframe in this scenario.    
	\end{itemize}
	In this way we have described the full singularity structure for the inhomogenous EOS corresponding to $ f(\rho) = \rho^\alpha $ in an asymptotically safe cosmology. Crucially, we have shown that Type I and Type III singularities can occur both in finite times and also in infinite times while Type IV singularities can occur in a finite time in such a cosmology. We have also shown that there is a significant difference between the singularity structure in standard cosmology for such an EOS and that in our asymptotically safe cosmology. We have also shown that for this ansatz, a Type IV singularity is not possible.  We take a moment here to look at the status quo of energy conditions in the scenarios we have discussed so far. In particular, we would like to discuss about the following energy conditions \begin{equation}
	\rho \geq 0 \quad \rho \pm p \geq 0 \qquad \text{"dominant energy condition"}
	\end{equation}
	\begin{equation}
	\rho + p \geq 0 \qquad \text{"null energy condition"}
	\end{equation}
	\begin{equation}
	\rho \geq 0 \quad  \rho + p \geq 0 \qquad \text{"weak energy condition"}
	\end{equation}
	\begin{equation}
	\rho + 3p \geq 0 \quad \rho + p \geq 0 \qquad \text{"strong energy condition"}
	\end{equation}
	We would like to comment about their status in the different scenarios we have outlined so far as follows \begin{itemize}   
		\item For $\alpha > 0$ , $A > 0$ and $\rho \to \infty$ , all of the strong, weak, null and dominant energy conditions are violated. 
		\item For $\alpha > 0 $, $ A < 0$ and $\rho \to \infty$ , all energy conditions are satisfied except the dominant energy condition.
		\item For $\alpha > 0 $ and $\rho \to \infty$ , all energy conditions are satisfied except the strong energy condition and in this case it does not matter whether A has a positive or negative value. 
		\item For $ A > 0 $ and $\rho \to 1/2 $ , all energy conditions are violated and this result is independent of whether $\alpha$ has a positive or negative value.
		\item For $ \alpha < 0 $ , $ A < 0 $ and $ \rho \to 1/2 $, all energy conditions are satisfied except the dominant energy condition.
		\item For For $ \alpha > 0 $ , $ A < 0 $ and $ \rho \to 1/2 $, all energy conditions are satisfied and no condition is violated.
	\end{itemize} 
	
	\subsection{$f(\rho) = C(\rho_{o} - \rho)^{-\gamma} $}
	Let's now consider the case $ f(\rho) = C(\rho_{o} - \rho)^{-\gamma} $, where C is an arbitrary constant and $\gamma$ is another real constant but it is assumed to be positive,  which was considered in \cite{Nojiri:2005sx} to address type II singularities in the context of (2.11). For this case, the scale factor and the $\rho$-t relation in usual cosmology are \begin{equation}
	a \sim a_{o} \exp \left[ - \frac{(\rho_{o} - \rho)^{\gamma +1 } }{3 C (\gamma + 1)}  \right]
	\end{equation}
	\begin{equation}
	t \sim t_{o} - \frac{(\rho_{o} - \rho)^{\gamma +1 } }{3 \kappa C \sqrt{3 \rho_{o}}  (\gamma + 1)}
	\end{equation}
	where $\kappa$ is again a constant. The scale factor and the $\rho$ -t relation in the asymptotically safe scenario are given by \begin{equation}
	a(\rho) = a_{o} \exp \left[ \frac{(\rho_{o} - \rho)^{\gamma +1 }  \left( 2 \rho_{o} - {_{2} F_{1}} (1,1+\gamma;2+\gamma;1 - \frac{\rho}{\rho_{o}})  \right)}{6 (\rho_{o}(1+\gamma))}  \right]
	\end{equation}
	\begin{equation}
	t - t_{0} = \frac{1}{9 C \rho_{o} N \rho^{1/4} } \left[6 (\rho_{o} - \rho)^{\gamma + 1} -2 \rho^{\gamma + 1} (3+2 \rho_{o} + 4 \gamma) {_{2} F_{1}} \left(\frac{3}{4},-\gamma;\frac{7}{4} ; \frac{\rho}{\rho_{o}}\right) \right]
	\end{equation}
	where $ {_{2} F_{1}} $ are the hypergeometric functions which are given by the hypergeometric series \begin{equation}
	{_{2} F_{1}} (a,b;c;z) = \sum_{n=0}^{\infty} \frac{(a)_{n} (b)_{n}}{(c)_{n}} \frac{z^n}{n !} =   1 + \frac{ab}{c} \frac{z}{1 !} + \frac{a (a+1) b (b+1 )}{c(c+1)} \frac{z^2}{2 !} +... 
	\end{equation}
	where $(q)_{n}$ are the Pochhammer symbols defined as, $$ (q)_{n} = \left\{
	\begin{array}{ll}
	1 & \quad n= 0 \\
	q(q+1)...(q+n-1) & \quad n > 0
	\end{array}
	\right.$$
	Its firstly very interesting to see that the simple forms of the scale factor and $\rho$-t relation (3.20-3.21) are changed quite substantially in this scenario, more so with the entry of an exotic hypergeometric function. But what is eye catching here is that the ansatz $ f(\rho) = C (\rho_{o} -\rho)^{\gamma} $ still provides Type II singularities in the same way as it does in usual cosmology. In the case of the scale factor(3.22), as $ \rho \to \rho_{o}$ the scale factor still remains finite as the third argument of the hypergeometric function, $ 2+\gamma$ will always be positive and so the function will not blow up. Thereby in this case we can have diverging pressure with finite scale factor. For (3.23), the second argument of the Hypergeometric function, $-\gamma$ , will always be negative and hence the Hypergeometric function will terminate and will be finite. Hence, even for $ \rho \to \rho_{o} $, time t can still be finite. So we can successfully have type II singularities even in an asymptotically safe cosmology with this particular ansatz.
	\\
	\\
	Furthermore it is interesting to look at the Hubble parameter in this scenario, which for the scale factor (3.22) and this ansatz is given by \begin{equation}
	H = \frac{6 (\gamma +1) C \rho_{o} M^{1/4} \sqrt[4]{\rho} (\rho_{o}-\rho)^{-\gamma -1}}{2 \rho_{o}-\, _2F_1\left(1,\gamma
		+1;\gamma +2;1-\frac{\rho}{\rho_{o}}\right)}
	\end{equation} 
	For this form of the Hubble paramter, we can write its second time derivative as \begin{multline}
	\ddot{H} = \frac{27 (\gamma +1) C^3 \rho_{o} M^{3/4} \rho ^{3/4} (k-\rho )^{-3 (\gamma +1)} \left((k-\rho ) ,
		_2F_1\left(1,\gamma +1;\gamma +2;1-\frac{\rho }{\rho_{o}}\right) \left((2 \rho_{o}\rho + \rho_{o}) \right) \right)}{(1-2 \rho )^3 \left(2 \rho_{o} -\, _2F_1\left(1,\gamma +1;\gamma
		+2;1-\frac{\rho }{\rho_{o}}\right)\right){}^3} \\ +  \frac{2 \rho_{o} (3 (\gamma
		+1)-2 \rho_{o}(2 \rho +1) +2 \rho  (-8 \gamma +10 \gamma  \rho +8 \rho -5))}{(1-2 \rho )^3 \left(2 \rho_{o} -\, _2F_1\left(1,\gamma +1;\gamma
		+2;1-\frac{\rho }{\rho_{o}}\right)\right){}^3}  \\ + \frac{4 \rho_{o}^2 (4 (\gamma
		+1)^2+ \rho_{o}^2 (2 \rho +1)-3 (\gamma +1) k)}{(1-2 \rho )^3 \left(2 \rho_{o} -\, _2F_1\left(1,\gamma +1;\gamma
		+2;1-\frac{\rho }{\rho_{o}}\right)\right){}^3} \\ + \frac{4 \rho_{o}^2 \left(-2 \rho_{o} \rho  (-7 \gamma +8 (\gamma +1) \rho -5)-2 (4 \gamma +3)^2
		\rho ^3+\left(48 \gamma ^2+82 \gamma +37\right) \rho ^2-3 (\gamma +1) (8 \gamma +7)
		\rho \right)}{(1-2 \rho )^3 \left(2 \rho_{o} -\, _2F_1\left(1,\gamma +1;\gamma
		+2;1-\frac{\rho }{\rho_{o}}\right)\right){}^3} \\ + \frac{2 \rho_{o} (3 (\gamma
		+1)-2 \rho_{o}(2 \rho +1) +2 \rho  (-8 \gamma +10 \gamma  \rho +8 \rho -5))}{(1-2 \rho )^3 \left(2 \rho_{o} -\, _2F_1\left(1,\gamma +1;\gamma
		+2;1-\frac{\rho }{\rho_{o}}\right)\right){}^3} \\ - \frac{\rho  (2 \gamma -4
		\gamma  \rho -2 \rho -1) \, _2F_1\left(1,\gamma +1;\gamma +2;1-\frac{\rho }{\rho_{o}}\right)}{(1-2 \rho )^3 \left(2 \rho_{o} -\, _2F_1\left(1,\gamma +1;\gamma
		+2;1-\frac{\rho }{\rho_{o}}\right)\right){}^3}
	\end{multline}
	While the expression above looks humongous, its very easy to see that there is a finite value for which $\ddot{H}$ diverges. Namely for $ \rho \to \frac{1}{2} $ , $\ddot{H} \to \infty $ while the scale factor (3.22) and the pressure given by the ansatz remain finite. Hence as only the higher derivatives of H diverge while $\rho$, p and a remain finite, we see that we can have type IV singularity for this ansatz as well. Furthermore, for $ \rho \to \frac{1}{2} $ the cosmic time t given by (3.23) remains finite as well and so the type IV singularity can occur in finite time. We would also like to suggest the reader to have a look at the appendix for a short discussion of other forms of EOS in this scenario. Finally we make comments on the status of energy conditions for the various scenarios we have discussed so far as follows \begin{itemize}
		\item For $ C > 0$ and $ \rho \to 0 $ , the null, weak and dominant energy conditions are satisfied but the strong energy condition is violated. 
		\item For $ C < 0$ and $ \rho \to 0 $, the null, weak and strong energy conditions are satisfied but the dominant energy condition is violated. 
		\item For $ C > 0 $ and $ \rho \to 1/2$, all energy conditions are violated. 
		\item For $ C < 0$ and $ \rho \to 1/2 $, all energy conditions are satisfied except the strong energy condition.
	\end{itemize}
	\section{Possible singularity removal methods}
	In this section we would like to address some possible methods of singularity removal which have been previously discussed in the literature as well. In particular, we would like address singularity removal by the account of conformal anomaly effects and $f(R)$ gravity realizations. The effect of quantum backreaction of conformal matter around Type I, Type II and Type III singularities were taken into consideration in \cite{Nojiri:2005sx,Nojiri:2004ip,Nojiri:2000kz}. In these cases, the curvature of the universe becomes
	large around the singularity time $ t= t_{s} $, although the
	scale factor a is finite for type II and III singularities.
	Since quantum corrections usually contain the powers
	of the curvature or higher derivative terms, such correction terms are important near the singularity. At this point, it becomes important to add a bit of context about what conformal anomalies are and how they are usually perceived in high energy physics. It is fair to assume that there are many matter fields during inflation in the early universe because the Standard Model of particle physics has almost 100 fields, and this number may increase by two if the Standard Model is contained in a supersymmetric theory. Although the behaviour of these (massless) matter fields—scalars, the Dirac spinors, and vectors in curved space-time—is conformal invariant, some divergences are observed because of the presence of the one-loop vacuum contributions. In the renormalized action, some counterterms are required to break the matter action's conformal invariance in order to cancel the poles of the divergence component. From the classical point of view, the trace of the energy momentum tensor in a conformally invariant theory is null. But  renormalization procedures can lead to the trace of an anomalous energy momentum tensor, which is the so-called quantum anomaly or the conformal anomaly(we would recommend the reader \cite{deser1976non,duff1994twenty,birrell1984quantum,Bamba:2014jia} for more details on conformal anomaly effects). The conformal anomaly we have described be considered to have the following form \cite{Nojiri:2005sx} \begin{equation}
	T_{A} = b \left( F + \frac{2}{3} \Box R \right) + b^{\prime} G + b^{\prime \prime} \Box R
	\end{equation} 
	where $T_{A}$ is the trace of the stress-energy tensor, F is the square of the 4d Weyl tensor and G is a Gauss-Bonet curvature invariant, which are given by \begin{equation}
	F = (1/3) R^2 - 2 R_{i j} R^{i j} + R_{i j k l} R^{i j k l }
	\end{equation} 
	\begin{equation}
	G = R^2 - 4 R_{i j} R^{i j } + R_{i j k l} R^{i j k l}
	\end{equation}
	b and $ b^{\prime} $ on the other hand are given by \begin{equation}
	b = \frac{N + 6 N_{1/2} + 12 N_{1} + 611 N_{2} - 8 N_{HD}}{120 (4 \pi)^2 }
	\end{equation} 
	\begin{equation}
	b^{\prime} = - \frac{N + 11 N_{1/2} + 62 N_{1} + 1411 N_{2} - 28 N_{HD} }{360 (4 \pi)^2 }
	\end{equation}
	with N scalar, $ N_{1/2} $ spinor , $ N_{1} $ vector fields  , $ N_{2} $ (= 0 or 1 ) gravitons and $ N_{HD} $ being higher derivative conformal scalars. For usual matter, $ b > 0 $ and $ b^{\prime} < 0 $ except for higher derivative conformal scalars while $ b^{\prime \prime} $ can be arbitrary. Quantum effects due to the conformal anomaly act as a fluid with energy density $\rho_{A}$ and pressure
	$ p_{A} $. The total energy density is $ \rho_{tot} = \rho + \rho_{A} $ . The conformal anomaly, also known as the trace anomaly, can be given by the trace of the fluid stress-energy tensor \begin{equation}
	T_{A} = - \rho_{A} + 3 p_{A}
	\end{equation} 
	The conformal anomaly corrected pressure and energy densities still obey the continuity equation (2.10) and using that, we can write \begin{equation}
	T_{A} = - 4 \rho_{A} - \frac{\dot{\rho_{A}}}{H} \left( \frac{1}{2 \rho_{A}} - 1 \right) 
	\end{equation}  
	We can immediately contrast this form of the conformal anomaly with the one obtained for standard cosmology in \cite{Nojiri:2005sx} \begin{equation}
	T_{A} = - 4 \rho_{A} - \frac{\dot{\rho_{A}}}{H}
	\end{equation}
	And in standard cosmology using (4.8) one can express $\rho_{A}$ as an integral in terms of $T_{A} $ as \begin{equation}
	\rho_{A} = - \frac{1}{a^4} \int a^4 H T_{A} dt
	\end{equation} 
	Furthermore $T_{A} $ can be expressed in terms of the Hubble parameter as \begin{equation}
	T_{A} = -12 b \dot{H}^{2} + 24 b^{\prime} (-\dot{H}^2 + H^{2} \dot{H} + H^{4} ) - (4b + 6 b^{\prime \prime} ) (H^{(3)} + 7 H \ddot{H} + 4 \dot{H}^2 + 12 H^{2} \dot{H})
	\end{equation}
	And using this, one can have an expression for $ \rho_{A} $ taking into account conformal anomaly effects near the singularity. But obtaining a corresponding integral for $\rho_{A}$ for the equation (4.7) is not possible in the same way as in (4.10). Hence its not feasible to address a possible removal of Type I- Type III singularities using conformal anomaly effects in this asymptotically safe cosmology. Furthermore, it was shown in \cite{Nojiri:2008fk} that $f(R)$ gravity effects can also facilitate in singularity removal. To understand whether $f(R)$ effects can do similarly for this cosmology, we modify the action (2.2) for the $f(R)$ gravity case as \begin{equation}
	\Gamma_{k} = \frac{1}{16 \pi G_{k}} \int d^4 x \sqrt{-g} \left[f(R) + 2 \Lambda_{k}\right] 
	\end{equation}
	Although the continuity equation remains the same as (2.10) for the $f(R)$ case, the Friedmann equation changes substantially. While we can consider many complex $f(R)$ forms, for now lets consider the simplest form $ f(R) = R + f_{2} R^2 $. In this scenario, the Friedmann equation, takes the form \cite{Bonanno:2010bt} \begin{equation}
	3 H^2 + 18 f_{2} (2 H \ddot{H} - \dot{H}^2 + 6 H^2 \dot{H}) = 8 \pi G_{k} \rho + \Lambda_{k}
	\end{equation}
	The simplest $f(R)$ modification for this asymptotically safe cosmology, as shown above, is still very complex.In fact, trying to ascertain the singularity structure for various  inhomogeneous EOS as we have done in section 3 for the usual action (2.2) is not looking feasible in this case and hence considering any potential singularity removals through $f(R)$ gravity as well does not look like a very optimistic prospect. So we conclude that the chances singularity removals in an inhomogenous EOS asymptotically safe cosmology through conformal anomaly and $f(R)$ gravity effects look very bleak.  
	\section{Concluding remarks}
	In this paper we have considered the formation of cosmological singularities in an asymptotically safe cosmology with inhomogenous equations of state(EOS). While such explorations were previously done in various modified gravity and quantum gravitationally inspired cosmologies (besides standard general relativistic cosmology), no such an endeavour has been taken up till now in the realms of renormalization group approaches to cosmology and we have hence took up on this journey. We briefly discussed about the basics of asymptotic safety and how it could be applied to gravity and cosmology. We also showed that such a scenario permits a transition between phantom and quintessence forms of universal evolution After this we discussed a particular form of asymptotically safe cosmology where the cut off scale is proportional to the Hubble parameter and we presented the cosmological basics like the Friedmann equation of such a cosmology. We then considered an inhomogeneous form of EOS, namely $ p = - \rho - f(\rho) $, which has been well motivated physically from various cosmological approaches. We then considered two different ansatz for $ f(\rho) $ and showed that one can have Type I - Type IV singularities in such a cosmology, where we demonstrated plenty of cases where the singularities can occur both in finite and infinite time. The conditions in which these singularities occur is significantly different than how they occur in the standard cosmology, with the formulations being even more involved. Interestingly this difference comes about without the need of any free parameters in the cosmological models, which is usually what one observes when one goes for the same pursuit via various modified gravity/ QG inspired cosmological approaches. Furthermore, we showed that usual singularity removal methods like conformal anomaly effects and $f(R)$ gravity effects do not offer much hope for singularity removal in this cosmology. In conclusion we have discussed in detail, for the first time, how cosmological singularities can form in asymptotically safe cosmologies with various inhomogeneous EOS and we hope that our work will propel more investigations towards singularities in such cosmologies. We would also like to mention something with regards to the novelty of the work here. One may raise the point that in recent years a lot of work has been focused on singularity formation in various cosmological settings and this work is yet another one in that fray. But what separates our work from others and provides its novelty is the paradigm we have considered. While other approaches since the initial work of Nojiri and Odintsov have considered such formations in modified gravity / quantum gravitational cosmologies which are paradigms where the background dynamics as a whole changes with a lot of free parameters ( like the brane tension for RS-II cosmologies etc. ) , the asymptotically safe cosmology we have considered does not completely changes the setup at the background level and nor does it rely on additional free parameters in order for it to give out a singularity formation status quo adequately different from that in usual GR for a similar scenario. It is interesting in the sense that while the background cosmology remains essentially like GR, the runnings of both the Newtonian coupling and the cosmological constant, alongside a modified continuity equation, give out a very fascinating singularity structure with the conditions for singularity formation vastly different than what it is for a usual GR based cosmology.
	
	\section{Acknowledgements}
	The work by MK has been supported by the grant of the Russian Science Foundation No-18-12-00213-P https://rscf.ru/project/18-12-00213/ and performed in Southern Federal University. The authors would also like to wholeheartedly thank Sergei Odintsov for helpful discussions with him on various aspects related to the work in this paper. We would also like to thank the reviewer of the paper for  constructive comments on the work.
	
	\section{Appendix A : Possible alternative EOS forms}
	In passing, we would also like to entertain the possibility of other forms of inhomogenous EOS. More exotic equations of states have also been discussed previously \cite{Nojiri:2005sr} , like \begin{equation}
	p = -\rho  + f(\rho) + G(H) 
	\end{equation} Let's consider a first example of this EOS as \begin{equation}
	p = w_{o} \rho + w_{1} H^{4}
	\end{equation} where $ w_{o} $ is the EOS parameter in the simple homogeneous case. Using the Friedmann equation (9), one can write \begin{equation}
	p = \left(w_{o} + w_{1} N \right) \rho
	\end{equation} Hence, we see that the EOS parameter w has shifted effectively as \begin{equation}
	w \to w_{eff} = w_{o} + N w_{1}
	\end{equation} 
	So we see that even if $ w_{o} < -1 $, as long as $ w_{eff} > -1 $ a big rip singularity does not occur. From the other side we
	can start with quintessence value of $ w_{o} $ and the inhomogeneous EOS (56) with sufficiently negative $ w_{1} $ will brings the
	cosmological evolution towards a phantom era. Here we see that the transition behaviour of EOS in asymptotically safe cosmology mimics of what it does in the standard cosmology as well \cite{Nojiri:2005sr}.
	\\
	\\
	Furthermore,we can also discuss an EOS of this form   \begin{equation}
	p  = - \rho  - A \rho^{\alpha} - B H^{4 \beta} 
	\end{equation}
	where $ f(\rho) = A \rho^{\alpha} $ and $ G(H) = B H^{4 \beta} $ . This EOS can reduced to the form of (2.11) as well by considering the Friedmann equation (2.9). thereafter which it becomes \begin{equation}
	p = - \rho - A \rho^{\alpha} - B^{\prime} \rho^{\beta} 
	\end{equation}
	We can write the scale factor and the $ \rho - t $ relation in this case as \begin{equation}
	a = a_{o} \exp \left(\frac{\rho^{-\beta} \left((1-\beta) \, _2F_1\left(1,-\frac{\beta}{\alpha-\beta};\frac{a}{\beta-\alpha}+2;-\frac{A \rho^{\alpha-\beta}}{B}\right)+2 \beta \rho \, _2F_1\left(1,\frac{\beta-1}{\beta-\alpha};\frac{\alpha-1}{\beta-\alpha}+2;-\frac{A \rho^{\alpha-\beta}}{B}\right)\right)}{6 (\beta-1) \beta
		B} \right)
	\end{equation}
	\\
	\begin{multline}
	t - t_{o} = \frac{\rho^{-\beta-\frac{1}{4}} \left(8 (4 \beta \rho+\rho) \, _2F_1\left(1,\frac{3-4 \beta}{4 a-4 \beta};\frac{\alpha-2 \beta+\frac{3}{4}}{\alpha-\beta};-\frac{A \rho^{\alpha-\beta}}{B}\right)\right)}{6 (4 \beta-3) (4 \beta+1) B M^{1/4} } + \\ \frac{4 (3-4 \beta) \, _2F_1\left(1,-\frac{\beta+\frac{1}{4}}{\alpha-\beta};\frac{\alpha-2
			\beta-\frac{1}{4}}{\alpha-\beta};-\frac{A \rho^{\alpha-\beta}}{B}\right)}{6 (4 \beta-3) (4 \beta+1) B M^{1/4} }
	\end{multline}
	While the expressions above look very interesting and would make up for a worthwhile analysis, we would not be pursuing this endeavour here as we have already discussed the existence of Type I-IV singularities in this scenario for two different ansatz both in finite and infinite time. Hence this exploration will be suitable for another time.
	
	\bibliography{JSPJMJasymp.bib}
	
	\bibliographystyle{unsrt}

\end{document}